\documentclass[12pt]{article}
\usepackage{graphicx}
\graphicspath{{pictures/}}
\DeclareGraphicsExtensions{.jpg}

\begin{document}

\title{\bf Sizes and Distances in High Energy Physics \footnote{The talk at the XXIII International Baldin Seminar on High Energy Physics Problems "Relativistic Nuclear Physics and Quantum Chromodynamics", September 19-24, 2016, Dubna, Russia. }}

\author{V.A.~Petrov \\
Institute for High Energy Physics,\\
``NRC Kurchatov Institute'', Protvino, RF}

\date{}

\maketitle

\begin{abstract}
This is a critical discussion of physical relevance of some space-time 
characteristics which  are in current use in high energy physics.
\end{abstract}

\section{Introduction}
High-energy physics is in many ways a synonymous of relativistic physics.
The latter circumstance introduces, as is well known, some peculiarities when one deals with coordinates, lengths, time intervals.
The theoretical and interpretational toolkit of high-energy physics  is dominated by energy-momentum characteristics while the spatio-temporal ones are directly unobservable and thus comparatively rare. Nonetheless, the latter are persistently present both in literature and physical discourses and sometimes play quite an important heuristic role. So Heisenberg's attempt in his S-matrix paradigm, developed later under the guise of the "analytic S-matrix approach" by G. Chew, to exorcise these entities from the particle physics as unobservable ones seems now failed.
However, spatio-temporal parameters often aren't clearly defined in quantum-field theoretical terms. Thereof, the need to at least reveal and clarify their true meaning.
In this talk I'll consider the following characteristics: charge radius, interaction range, interaction time and the size of the like-charge pion source.

\section{Charge Radius}

In PDG volumes one can find values of  "charge radii" of hadrons. E. g. the proton charge radius is presented in two hypostasis: $ r_{charge} (proton) = 0.84087\pm 0.000039 fm $ ($ \mu p $ Lamb shift)
and $ r_{charge} (proton) = 0.8775 \pm 0.0051 fm $ ( $ ep $ CODATA value).

How these radii are defined?
In non-relativistic quantum mechanics when considering , say, the scattering of an electron off a composite system(atom) the relationship between 
the average radius of the scatterer and the derivative of its form factor $F(\textbf{q}^{2})  $ 
is ( with $ \textbf{q} $ the transferred 3-momentum) 
\begin{center}
$$\langle\textbf{ r}^{2} \rangle = - 6dF(\textbf{q}^{2})/d\textbf{q}^{2} |_{\textbf{q}^{2} = 0}   \eqno{(1)} $$
\end{center}
where 
\begin{center}
$$ F(\textbf{q}^{2})= \int d\textbf{r} e^{i\textbf{qr}} \rho (\textbf{r}). \eqno{(2)} $$.
\end{center}
Here the average charge density is of the form \begin{center}
$$\rho(\textbf{r})=\sum_{k=1}^{N} e_{k}\rho_{k}(\textbf{r}) $$
\end{center}
where
\begin{center}
$$ \rho_{k}(\textbf{r})= \int \prod_{j\neq k} d\textbf{r}_{j} |\Psi( \lbrace \textbf{r}_{j}\rbrace,  \textbf{r}_{k}=\textbf{r})|^{2}. \eqno{(3)}$$
\end{center}
So we get
\begin{center}
$$ \langle \textbf{r}^{2} \rangle =\sum_{k=1}^{N} e_{k} N_{k}\langle \textbf{r}^{2}\rangle _{k} . $$
\end{center}
where $N_{k}  $ is the number of constituents with charge $ e_{k} $ 
and 
$$\sum_{k=1}^{N} e_{k} N_{k}$$ is the total charge of the system   ,   
$$ \sum_{k=1}^{N} N_{k} = N  $$ 
while $ \langle \textbf{r}^{2}\rangle _{k}  $  stands for the average square of the distance of the k-th constituent from the center of mass of the system which is implicitly fixed in Eq.(3)with the condition $$ \delta (\sum_{1}^{N} \textbf{r}_{i} m_{i} /\sum_{1}^{N} m_{i} ).
$$ 
From this expression it is evident that the "charge radius" generally doesn't give 
a clear idea of the geometrical extent of the system which is closer to the expression of the type 
\begin{center}
$$ \langle \textbf{r}^{2} \rangle = \frac{1}{N}\sum_{k=1}^{N} \langle \textbf{r}^{2}\rangle _{k} . $$
\end{center}
The wave function is a solution of the Schr$ \ddot{o} $dinger equation with an appropriate instantaneous potential. 
When coming to the relativistic domain the use of the instantaneous potential becomes problematic because of the retardation effects.
Nonetheless, Eq.(1)is being actively used for extracting from the data the "charge radius" of, say, the proton with further placing the result into the PDG Reviews as an important physical characteristic of the proton charge distribution.
To avoid as much as possible the model dependence we first consider the formula (1) in the context of the general quantum-field theoretic expression via the Bogoliubov-LSZ reduction formalism. For simplicity but without loss of generality we consider the pion form factor $ F(q^{2}) $ defined as follows
\begin{center}
$  \langle p+q \vert J^{\mu}(0) \vert p \rangle = (2p^{\mu} + q^{\mu})F(q^{2}) $
\end{center}
where $ J^{\mu}(x) $ is the electromagnetic current operator and $ \vert p \rangle $ is a charged pion state.The function $ F(q^{2}) $ is hermitian analytic in the complex $ q^{2} $ plane with a cut $ [4m_{\pi}^{2},+\infty] $.
In terms of the 4-momentum transfer $ q^{2} =q_{0}^{2} -\textbf{q}^{2}  $ 
the "charge radius" of the pion is taken as 

$$\langle \textbf{r}^{2} \rangle = 6dF({q}^{2})/d{q}^{2} |_{{q}^{2} = 0}   \eqno{(1^{*})} $$
The only known model-independent way to introduce space-time coordinates is the use of the reduction formulas connecting Green functions ( and hence scattering amplitudes) in momentum and configuration space.
Taking use of the reduction formalism [1] we get
\begin{center}
$ F({q}^{2}) = \frac{2p_{\mu}}{4m_{\pi}^{2}- {q}^{2}}\int d^{4}x e^{iqx} \langle 0\vert \frac{\delta J^{\mu}(x)}{\delta \phi^{+}(0)}\vert p \rangle$.
\end{center}
Variation derivative is taken over the pion out-field.
Up to a finite sum of quasi-local operators 
\begin{center}
$ \frac{\delta J^{\mu}(x)}{\delta\varphi^{+}(0)}= i\theta (-x)[J^{\mu}(x),I^{+}_{\pi}(0)]. $
\end{center}
Here $I^{+}_{\pi}(x) = i\frac{\delta S}{\delta \varphi^{+}_{\pi}(x)}S^{+} $ is the pion density operator.
For definiteness let us take the laboratory frame where $ \textbf{p} =0 $.
We get 
\begin{center}
$ F({q}^{2}) = \frac{2m_{\pi}}{4m_{\pi}^{2}- {q}^{2}}\int d^{4}x \exp [ -i\frac{q^{2}}{m_{\pi}} -i(\textbf{xn})\sqrt{-q^{2}(1-q^{2}/m_{\pi}^{2})}] \langle 0\vert \frac{\delta J^{0}(x)}{\delta \phi^{+}(0)}\vert \textbf{p}=0 \rangle$.
\end{center}
Now, if we apply to this representation  formula $(1^{*})$ for the "charge radius"
we have
$$\langle \textbf{r}^{2} \rangle = \int d\textbf{r}\textbf{r}^{2}\rho_{L}(\textbf{r}) \eqno{(4)} $$
where
$$\rho_{L}(\textbf{r}) = \frac{1}{2m_{\pi}}\int dx^{0}\langle 0\vert \frac{\delta J^{0}(x^{0},\textbf{r} )}{\delta \phi^{+}(0,\textbf{0})}\vert \textbf{p}=0 \rangle \eqno{(5)} $$
is to have the meaning of the charge density inside the pion in rest.

When looking at Eq.(5) we notice that the distance $ |\textbf{r}| $ relates points taken at different times $ x^{0}, 0 $ and so the profile of the supposed charge distribution $ \rho_{L}(\textbf{r}) $ doesn't give us an instantaneous snapshot of the charge distribution inside the pion but rather something smeared in time.
One can prove (with use e.g. of the JLD representation for causal commutators) that in the non-relativistic limit
\begin{center}
$ c\frac{1}{2m_{\pi}}\langle 0\vert \frac{\delta J^{0}(ct,\textbf{r} )}{\delta \phi^{+}(0,\textbf{0})}\vert \textbf{p}=0 \rangle |_{c \longrightarrow \infty} = \delta (t) \Phi (\textbf{r}) $
\end{center}
so we recover a NR quantum-mechanical expression like Eq.(3).
On the other hand, one can argue that non-simultaneity in the definition of the charge radius ( or a particle size) can be taken into account if to assume that arising uncertainty is given by the "retardation time"  $ \sim \left\langle r\right\rangle/c $ and may induce an uncertainty comparable with the very radius in question .

\section{ On "physical" proton radius}

With all reservations stemming from the previous Section let us consider the charge radius of the nucleon from the point of its use in modelling practice. Quite often the value of the charge radius of the proton is taken in models of high-energy scattering to account for the size of the colliding nucleons. The fact of the matter is that the charge distribution may generally differ very much from the "matter distribution" due to the presence of the constituent charges making the distribution generally non positive defined. E.g. the square of the charge radius of the neutron is \textbf{negative}, as follows from the definition (1*) and experimental data, and hence doesn't give us much to say about the neutron physical size. To make the point more concrete it is useful to express the charge radii of the proton and neutron in terms of the valence quarks.
With account of the isotopic symmetry it is not difficult to obtain the following expression for the nucleon \textbf{physical} size

\begin{center}
$  \left\langle r^{2}\right\rangle (nucleon) = r_{charge}^{2}(proton) + r_{charge}^{2}(neutron)$
\end{center}
Let's take for definiteness the PDG values $ r_{charge} (proton) = 0.707 fm^{2} $ ($ \mu p $ Lamb shift) and $ r_{charge}^{2} (neutron) = - 0.116 fm^{2} $. 

Thereof we get for the physical radius of the nucleon
\begin{center}
$  \sqrt{\left\langle r^{2}\right\rangle (nucleon)} = 0.769 fm $ 
\end{center}
to compare with $  r_{charge} (proton) = 0.841 fm  $.
At first sight the difference can seem insignificant but in cases where the energy dependence is logarithmic such difference leads to significant differences in the estimates of the relevant energies \cite{2} .
\section{ On the interaction range}

The transverse interaction range $ \left\langle b^{2}\right\rangle_{tot}  $ ( with $ \textbf{b} $ the impact parameter)is defined from the forward logarithmic slope $ B(s) $

\begin{center}
 $B(s)\equiv \left[\frac{\frac{\partial}{\partial t}\left(\frac{d\sigma}{dt}\right)}{\frac{d\sigma}{dt}}\right]_{t=0} = \partial ln [{d\sigma/dt}]/\partial t\mid _{t=0}  = 2Re (\partial \ln (T( s, t))/ \partial t)\mid_{t=0} $
 \end{center} 
so that
\begin{center}
$$2B(s)\approx\frac{\int db^2\,b^2\,{\rm Im\,\tilde{T}}(s,b)}{\int db^2\,{\rm Im\,\tilde{T}}(s,b)}=<b^2>_{tot}\ \eqno(5)$$
\end{center}
where $\tilde{T}(s,b)= \frac{1}{16\pi s }\int T(s,t)J_0(b\sqrt{-t})dt$.

It is a peculiar feature of the unitarity that the range of the transverse area where any process (\textit{inelastic} included)  can happen is defined by the quantity describing \textit{elastic} scattering.

One can define the quantity $ <b^2>_{el} $ which seems to be more relevant for spatial description of elastic scattering only

\begin{center}
$ (\int db^2 b^2 |\tilde{T}(s,b)|^{2})/(\int db^2 |\tilde{T}(s,b)|^{2}) =<b^2>_{el}.$
\end{center}
However it depends on the scattering phase which is very problematic as an observed quantity [3]:
\begin{center}
$<b^2>_{el} = <(-t)B^2(s,t)+ 4(-t)[\frac{\partial \Phi(s,t)}{\partial t}]^2>.$ 
\end{center}

Experiments show a very slow growth of $<b^2>_{tot}  $:
\begin{center}
$ \sqrt{<b^2>_{tot}} (0.01 TeV @  U-70, IHEP)\approx 0.9 fm $
\end{center}
while 
\begin{center}
$ \sqrt{<b^2>_{tot}} (7 TeV  @ LHC, CERN)\approx 1.3 fm $
\end{center}
only.
If to take into account that physical values of the "valence" sizes of colliding nucleons  $ \sqrt{ \frac{2}{3}\left\langle r^{2}\right\rangle (nucleon)}  $ (seen in the impact parameter plane) are, as we have seen above, of order of 0.6 fm   we conclude that colliding nucleons hardly cease to overlap even at the LHC energies. So even at the LHC we are very far from the "asymptopia" which implies the region of energies where 

\begin{center}
$ \sqrt{<b^2>_{tot}} \gg \sqrt{ \frac{2}{3}\left\langle r^{2}\right\rangle (nucleon)} . $
\end{center} 
 
 Can one argue about the \textit{longitudinal} range as well? Yes but it's again related to the almost unobservable scattering phase:
 \begin{center}
 $\left\langle L^{*}\right\rangle  = 4p^{*} \left\langle (\partial \Phi(s,t)/ \partial t) \right\rangle  $
 \end{center}
 where asterisk means the cms values and $p^{*}$ is the cms value of colliding hadrons' momentum.
 Here we can only say that existing models give a very fast grow of $ \left\langle L^{*}\right\rangle  $ proportional to the cms energy $ \sqrt{s} $.
\section{Time evolution of space-time correlation functions in hadron collisions}
 
 Let's consider the forward scattering amplitude. It can be cast in the following form 
 \begin{center}
 $$ T(k,p) = \int d^{4}x \exp (ikx) \left\langle p \vert \frac{\delta I (-x/2)}{\delta \varphi (x/2)}|p\right\rangle . \eqno {(6)} $$ 
 \end{center}
 For the sake of simplicity let us take the simple Regge pole model high-energy behaviour of this amplitude 
 \begin{center}
 $ T(k,p)\approx (2pk)^{\alpha (0)}\beta(k^{2}) $
 \end{center}
 where we allow the projectile to be off-shell and consider the region $ 2pk\gg k^{2} $.
 
 With a sufficient degree of rigour it was shown [4] that the correllator representing the amplitude in the configuration space behaves as follows:
 
\begin{center}
$$ \left\langle p|\frac{\delta I (-x/2)}{\delta \varphi (x/2)}|p\right\rangle \approx (px)^{\alpha (0)}g(x^{2})\eqno{(7)} $$
\end{center}
at
\begin{center}
 $ (px) \rightarrow \infty ,  x^{2}  $ fixed.
\end{center}
Simple Regge-pole asymptotics can be generalized to more complicated cases. E.g. in the case of the functional saturation of the Froissart bound we would have

\begin{center}
$$ \left\langle p|\frac{\delta I (-x/2)}{\delta \varphi (x/2)}|p\right\rangle \approx (px)\ln ^{2} (px) g(x^{2}). \eqno{(8)} $$
\end{center}

In the laboratory frame and with account of the causality condition 
 ($\delta I( -x/2)/\delta \varphi(x/2) = 0$ at 
 $ x\bar{\in}   (x: x^{0} \geq  |\textbf{x}|)   $)
we see that the correlation function $ \left\langle p|\frac{\delta I (-x/2)}{\delta \varphi (x/2)}|p\right\rangle  $ grows with the time $ t $ growth as $ \sim t^{\alpha (0)} $ at ${\alpha (0)} > 0 $ or as $ \sim t \ln ^{2}t $ inside the upper light cone on any fixed hyperboloid $ ct = \sqrt{x^{2} + \textbf{x}^{2}} $ (this corresponds to large values of one of the light-cone variables  $ x^{\pm} $).
Such a behaviour of the correlation function is quite unusual because  correlation functions we deal with in, say,  statistical physics  ordinarily die off at large times and distances.

We here should also mention the estimate of the interaction time which is being made supposing that it can be given by an "essential" space-time region where the exponential in Eq.(6) doesn't strongly oscillate. With such an assumption  some authors ( see. e.g. Ref.[5])obtained that the "effective interaction time"  $ \Delta t_{eff} $ can be estimated (in the laboratory frame, e.g.)as
\begin{center}
$ \Delta t_{eff} \approx E/m^{2} .$
\end{center}

Accordingly the longitudinal distances can be estimated as 
$ L_{eff} \sim \Delta t_{eff}$ so that at the LHC  we would have the distances ( reduced to the cms frame), where the scattering takes place, as large as 1500 fm.
Sure, such an estimate doesn't take into account the correlator behaviour inside the " effective integration region" but Eqs.(6, 7) seem to support such a growth.
\section{On the size of the like-charge pion source}
Many years ago A. S. Goldhaber et al observed that the like charge pions significantly correlate in contrast to the unlike charge pions with this correlation seen in various correlation measures. With the obvious reason for like-charge pions to correlate due to the Bose-Einstein statistics the main aim of the further studies of these correlations was to extract from the data the spatial size of the region where these pions are being radiated from.
Generic designation of this correlation function is R(Q) where R can be defined in various ways dependent on the reference distribution exemplifying the uncorrelated samples. Fig.1 [5] represents a typical behaviour of R as function of $ Q = \sqrt{-(k_{1}-k_{2})^{2}}  = \sqrt{M^{2}_{\pi\pi} - 4m^{2}_{\pi}}$.

\begin{figure}[h!]
\centering

\includegraphics[width=13cm,clip]{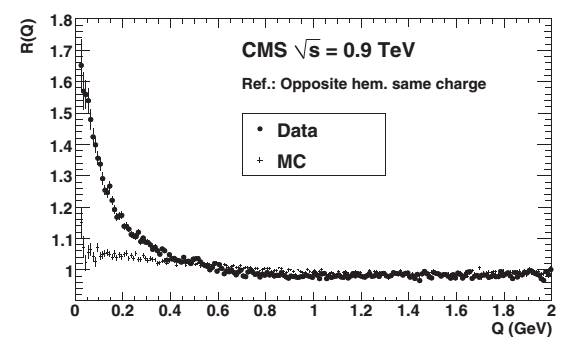}
\caption{Typical BE correlation function as measured by CMS.}
\label{fig-1}       
\end{figure}

Generic form (with insignificant deviations) of R is
\begin{center}
$$  R(Q) = 1 + \exp (-RQ) \eqno {( 8)} $$.
\end{center}
 And this is parameter R which is treated as the average size of the like-charge pion source.
Such an interpretation stems from the conviction that the two-particle distribution can be represented as a Fourier transform of the modulus squared of some function in the configuration space as it takes place for the form factor in NR QM (see Eq.(2)).
I didn't manage to find out a consistent theoretical derivation of this elegant formula (8). If to accept the premises used in literature for the spatial distribution of the "source size" $ r $ then, in order to get $ \exp (- QR) $, it should develop a singularity of the type
$1/(r^{2}+ R^{2}) $ 
which can be at odds with analyticity properties of Wightman functions in configuration space [1].
Without going into the details of very interesting findings concerning the energy and multiplicity dependence of the parameter R (or even two such parameters [6]) I would like just to show the integral representation of the two-pion distribution function in terms of quantum-field theoretical space-time correlators.
\begin{center}
$ \frac{dN(N-1)}{d^{3}k_{1}d^{3}k_{2}}  
\sim \int d^{4}X d^{4}\xi d^{4}\eta
 e^{iPX}e^{iq (\xi-\eta)}  \left\langle pp,in|\bar{T}[I (X+\xi/2) I (X - \xi/2)]
  T [ I(\eta/2) I (-\eta/2)]|pp,in \right\rangle$ 
\end{center}
where $ q = k_{1}- k_{2}, P = (k_{1}+ k_{2}) $ and $ I(x) $ is a pion density operator.
First of all we see that in configuration space the integrand is non-diagonal which complicates its probabilistic interpretation.
In any relevant channel the distance $ \xi $ refers to the 4-distance between the radiated pions. At the same time the momentum $ q $ related to the quantity Q is conjugated both to $ \xi $ and $ \eta $.  So it is difficult to relate the parameter R to the average source size. It remains to believe that we finally will eliminate these conceptual inconveniences though the problem seems quite a complicated from the formal theoretical viewpoints.
I have, nonetheless, to stress that  this statement in no way devalues intriguing regularities found in experimental and phenomenological studies for the parameter R. The problem is only what does it mean.

\section{Conclusion}

What was said in this talk is mostly problem statements related to the lack of sound reasons for the standard physical interpretation of some space-time quantities, as they are normally defined, which I believe deserve the attention of the audience and hopefully will motivate further discussions and help to find acceptable solutions.

\section{Acknowledgements}
I am grateful to the organizers of the XXIII International Baldin Seminar on High Energy Physics Problems for invitation, hospitality and very good organization of the Seminar.

\end{document}